\documentclass[aps,pra,twocolumn,superscriptaddress,letterpaper]{revtex4-1}
\usepackage{graphicx}
\usepackage{amsmath}
\usepackage{esint}
\usepackage{verbatim}
\usepackage{color}
\usepackage{SIunits}
\usepackage{hyperref}

\newcommand{\nbar}{\langle n \rangle}

\newcommand{\nbathp}{n_{\text{p}}}
\newcommand{\Tbathp}{T_{\text{p}}}
\newcommand{\Tbath}{T_{\text{b}}}
\newcommand{\Tf}{T_{\text{f}}}
\newcommand{\nfridge}{n_{\text{f}}}

\newcommand{\lambdacO}{\lambda_{\text{c}}}
\newcommand{\ncavO}{n_{\text{c}}}
\newcommand{\gzeroO}{g_{\text{0}}}

\newcommand{\kappai}{\kappa_{\text{i}}}
\newcommand{\kappae}{\kappa_{\text{e}}}
\newcommand{\gammaiO}{\gamma_{\text{i}}}
\newcommand{\gammapO}{\gamma_{\text{p}}}
\newcommand{\gammanotO}{\gamma_{\text{0}}}
\newcommand{\omegacO}{\omega_{\text{c}}}
\newcommand{\omegamO}{\omega_{\text{m}}}
\newcommand{\gammaOMO}{\gamma_{\text{OM}}}

\newcommand{\QmO}{Q_{\text{m}}}

\newcommand{\omegaS}{\omega_{\text{s}}}
\newcommand{\omegaLO}{\omega_{\text{LO}}}
\begin{document}

\title{Thermalization properties at mK temperatures of a nanoscale optomechanical resonator with acoustic-bandgap shield}

\author{Se\'{a}n M.\ Meenehan}
\thanks{These authors contributed equally to this work.}
\author{Justin D.\ Cohen}
\thanks{These authors contributed equally to this work.}
\affiliation{Institute for Quantum Information and Matter and Thomas J. Watson, Sr., Laboratory of Applied Physics, California Institute of Technology, Pasadena, CA 91125, USA}
%\affiliation{Institute for Quantum Information and Matter, California Institute of Technology, Pasadena, CA 91125, USA}
\author{Simon Gr\"oblacher}
\thanks{These authors contributed equally to this work.}
\affiliation{Institute for Quantum Information and Matter and Thomas J. Watson, Sr., Laboratory of Applied Physics, California Institute of Technology, Pasadena, CA 91125, USA}
%\affiliation{Institute for Quantum Information and Matter, California Institute of Technology, Pasadena, CA 91125, USA}
\affiliation{Vienna Center for Quantum Science and Technology (VCQ), Faculty of Physics, University of Vienna, A-1090 Wien, Austria}
\author{Jeff T.\ Hill}
\author{Amir H.\ Safavi-Naeini}
\affiliation{Institute for Quantum Information and Matter and Thomas J. Watson, Sr., Laboratory of Applied Physics, California Institute of Technology, Pasadena, CA 91125, USA}
%\affiliation{Institute for Quantum Information and Matter, California Institute of Technology, Pasadena, CA 91125, USA}
%\author{Jasper Chan}
%\affiliation{Thomas J. Watson, Sr., Laboratory of Applied Physics, California Institute of Technology, Pasadena, CA 91125, USA}
\author{Markus Aspelmeyer}
\affiliation{Vienna Center for Quantum Science and Technology (VCQ), Faculty of Physics, University of Vienna, A-1090 Wien, Austria}
\author{Oskar Painter}
\email{opainter@caltech.edu}
\affiliation{Institute for Quantum Information and Matter and Thomas J. Watson, Sr., Laboratory of Applied Physics, California Institute of Technology, Pasadena, CA 91125, USA}
%\affiliation{Institute for Quantum Information and Matter, California Institute of Technology, Pasadena, CA 91125, USA}
%\affiliation{Max Planck Institute for the Science of Light, D-91058 Erlangen, Germany}

\date{\today}
\begin{abstract}
Optical measurements of a nanoscale silicon optomechanical crystal cavity with a mechanical resonance frequency of $3.6$~GHz are performed at sub-kelvin temperatures. We infer optical-absorption-induced heating and damping of the mechanical resonator from measurements of phonon occupancy and motional sideband asymmetry. At the lowest probe power and lowest fridge temperature ($\Tf=10$~mK), the localized mechanical resonance is found to couple at a rate of $\gammaiO/2\pi=400$~Hz ($\QmO = 9\times 10^6$) to a thermal bath of temperature $\Tbath \approx 270$~mK. These measurements indicate that silicon optomechanical crystals cooled to millikelvin temperatures should be suitable for a variety of experiments involving coherent coupling between photons and phonons at the single quanta level. 
\end{abstract}
\maketitle

\emph{Introduction.~-}~The coupling of a mechanical object's motion to the electromagnetic field of a high finesse cavity forms the basis of various precision measurements~\cite{Braginsky1977}, from large-scale gravitational wave detection~\cite{LIGO2011} to microscale accelerometers~\cite{Krause2012}. Recent work utilizing both optical and microwave cavities coupled to mesoscopic mechanical resonators has shown the capability to prepare and detect such resonators close to their quantum ground-state of motion using radiation pressure back-action~\cite{Rocheleau2010,Teufel2011b,Chan2011,Verhagen2012}.  Optomechanical crystals (OMCs), in which bandgaps for both optical and mechanical waves can be introduced through patterning of a material, provide a means for strongly interacting nanomechanical resonators with near-infrared light~\cite{Eichenfield2009b}. Beyond the usual paradigm of cavity-optomechanics in which mechanical motion is confined to single objects, such as movable end-mirrors~\cite{Groeblacher2009a} and intra-cavity membranes~\cite{Thompson2008}, OMCs can be fashioned into planar circuits for photons and phonons, and arrays of optomechanical elements can be interconnected via optical and acoustic waveguides~\cite{Safavi-Naeini2010b}. Such coupled OMC arrays have been proposed as a way to realize quantum optomechanical memories~\cite{Chang2011}, nanomechanical circuits for continuous variable quantum information processing~\cite{Schmidt2012} and phononic quantum networks~\cite{Habraken2012}, and as a platform for engineering and studying quantum many-body physics of optomechanical meta-materials~\cite{Tomadin2012,Ludwig2013,Schmidt2013}.

The realization of optomechanical systems in the quantum regime is predicated upon the ability to limit thermal noise in the mechanics while simultaneously introducing large coherent coupling between optical and mechanical degrees of freedom. In this regard, laser back-action cooling has recently been employed in simple OMC cavity systems~\cite{Chan2011,ChanPhD} consisting of a one-dimensional (1D) nanobeam resonator surrounded by a two-dimensional (2D) phononic bandgap. In this work, we optically measure the properties of such an OMC cavity system in a helium dilution refrigerator down to base temperatures of $\Tf=10$~mK. 

%Owing to the extreme isolation provided by the phononic bandgap, even at the lowest probe powers ($2$~nW) the localized $3.6$~GHz mechanical resonance couples to an effective thermal bath of temperature $270$~mK, with relatively weak coupling to the external fridge bath. 

\begin{figure}[btp]
\begin{center}
\includegraphics[width=\columnwidth]{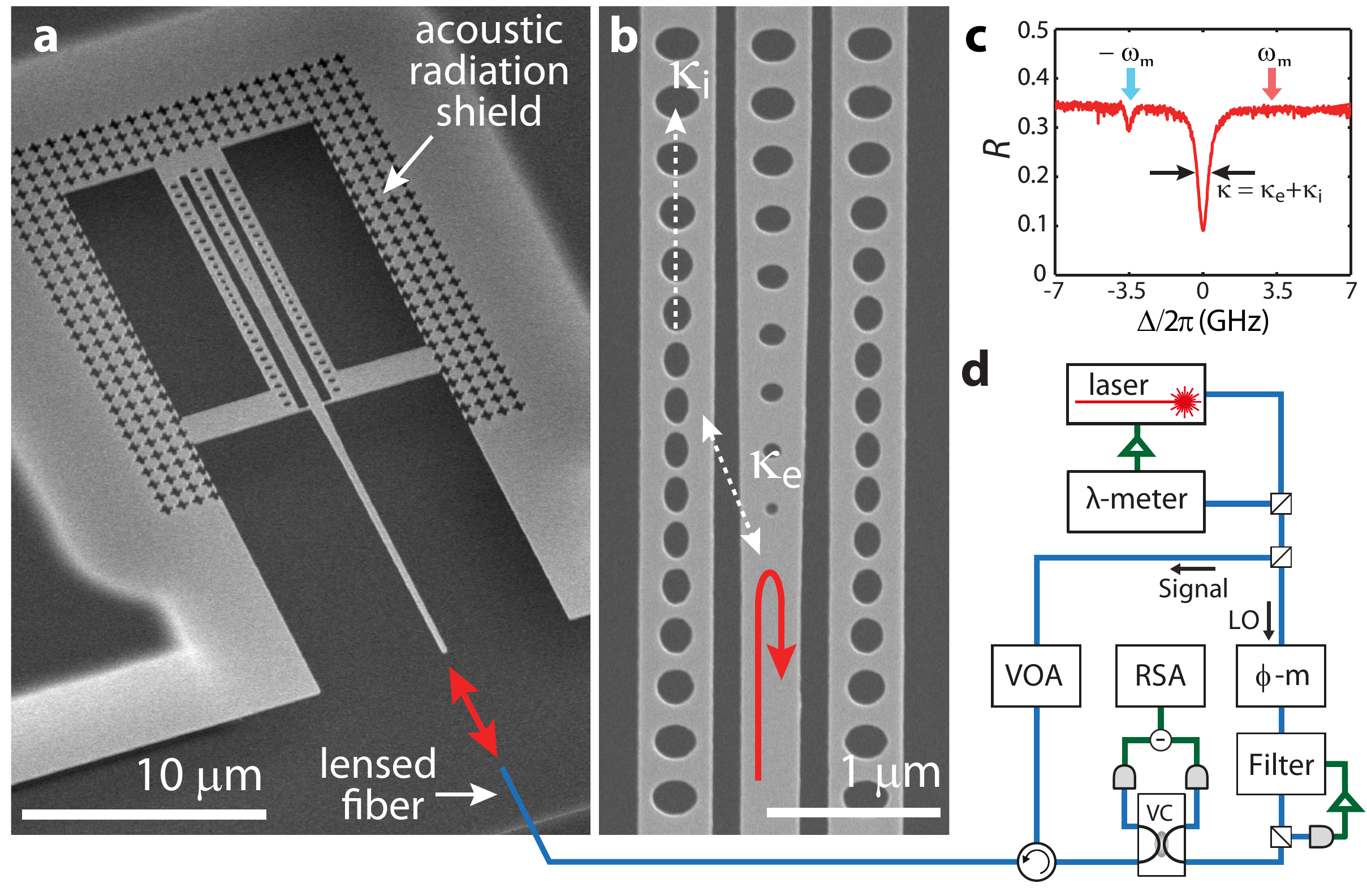}
\caption{\textbf{Experimental setup.} \textbf{a}, SEM image of the OMC device, with acoustic radiation shield and end-fire fiber coupling indicated. \textbf{b}, Zoom in SEM image showing details of the waveguide--cavity coupling region, indicating the extrinsic cavity-waveguide coupling rate $\kappae$ and the intrinsic loss rate $\kappai$. \textbf{c}, Normalized optical cavity reflection spectrum ($R$). Detunings from resonance of $\Delta = \pm\omegamO/2\pi=\pm 3.6$~GHz are denoted by the red and blue arrows, respectively. \textbf{d}, Schematic of the fiber-based heterodyne receiver used to perform optical and mechanical spectroscopy of the OMC cavity. $\lambda$-meter:\ wavemeter, $\varphi$-m:\ electro-optic phase modulator, VOA:\ variable optical attenuator, VC:\ variable coupler, RSA:\ real-time spectrum analyzer.} \label{fig:setup}
\end{center} 
\end{figure}

The device studied here is formed from the top silicon (Si) device layer of a silicon-on-insulator (SOI) wafer using a combination of electron beam lithography, plasma etching, and wet etching (see also App.~\ref{sec:appB}). Figure~\ref{fig:setup}a shows a scanning electron microscope (SEM) image of a suspended device after processing. This device consists of two nanobeam OMC cavities, optically coupled to a common central waveguide (Fig.~\ref{fig:setup}b).  As described in Ref.~\cite{Chan2011}, the nanobeam cavities are patterned in such a way as to support an optical resonance in the $1550$~nm wavelength band and a 'breathing' mode mechanical resonance at $3.6$~GHz.  The highly localized optical and acoustic resonances couple strongly via radiation pressure, with a theoretical vacuum coupling rate of $\gzeroO/2\pi=870$~kHz, corresponding physically to the optical resonance shift due to zero-point fluctuations of the mechanical resonator.  Surrounding the waveguide and nanobeam structure on three sides is a 2D `cross' pattern~\cite{Safavi-Naeini2010b} which has a full phononic bandgap for all acoustic waves in the frequency range from $3-4$~GHz. The fourth side is left open, and a final set of etches is used to clear the buried oxide and silicon handle wafer so as to allow close approach of an optical fiber.  The SOI sample is mounted to the mixing chamber plate of a dilution refrigerator, and a set of position encoded 'slip-stick' stages are used to align an anti-reflection-coated tapered lensed fiber (beam waist=2.5~$\mu$m; focal distance=14~$\mu$m) to the coupling waveguide of a given device under test (see Fig.~\ref{fig:setup}a and App.~\ref{sec:appC}).  In order to aid efficient optical coupling, the Si waveguide is tapered down to a tip of width $225$~nm, providing mode-matching between fiber and waveguide~\cite{Almeida2003}.

%~\cite{Mitomi1994,Almeida2003,Cohen2013}

\begin{figure}[btp]
\begin{center}
\includegraphics[width=\columnwidth]{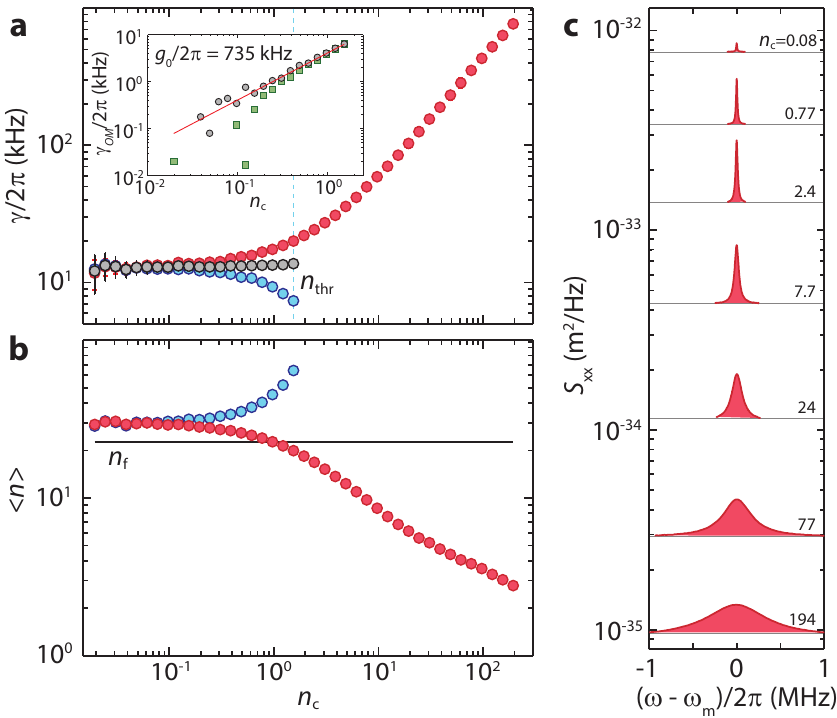}
\caption{\textbf{$T=4$~K data.} \textbf{a}, Measured mechanical linewidth $\gamma$ for $\Delta=\omegamO$ (red) and $-\omegamO$ (blue) at a fridge temperature of $\Tf=4$~K. The vertical blue dashed line indicates the threshold $\ncavO$ beyond which the mechanical resonance self-oscillates. Black circles indicate the values of $\gammaiO$ obtained by taking the average of the detuned data. The inset shows the optomechanical damping rate $\gammaOMO$ determined by two methods: in circles $\gammaOMO$ is extracted by subtracting $\gammaiO$ from the red-detuned $\gamma$, and in squares $\gammaOMO$ is found by determining cooperativity $C$ using the calibrated $\nbar$. A linear fit (red line) yields $\gzeroO/2\pi = 735$~kHz. \textbf{b}, Calibrated mechanical mode occupancy $\nbar$ versus intracavity photon number $\ncavO$. Blue and red circles are measured with probe laser detunings $\Delta=\pm\omegamO$, respectively. The mode occupancy $\nfridge$ corresponding to $\Tf=4$~K is indicated by a black solid line. \textbf{c}, Series of red-detuned NPSD for range of $\ncavO$.  Here the NPSD is plotted as $S_{\text{xx}} = x_{\text{zpf}}^2 S_{\text{bb}}$, where $x_{\text{zpf}}=4.1$~fm is the zero-point amplitude of the mechanical resonator.} \label{fig:4K}
\end{center} 
\end{figure}

Optical spectroscopy of a fiber-coupled device at a fridge temperature of $4$~K is shown in Fig.~\ref{fig:setup}c. In this measurement, the frequency of a narrow linewidth external-cavity diode laser is swept across the fundamental optical resonance of one of the OMC cavities centered at $\lambdacO = 1545$~nm. The input laser light is reflected from a photonic crystal mirror at the end of the Si waveguide, and when resonant, evanescently couples in to one or the other of the nanobeam cavities (the other nanobeam cavity coupled to this waveguide has a resonance several THz to the red). This geometry allows for the extrinsic coupling rate, $\kappae$, to the OMC cavities to be adjusted via the waveguide-cavity gap size, and eliminates optical loss in transmission past the cavity. Some of the light entering the cavity decays through intrinsic loss channels (i.e., ones we do not detect externally) at rate $\kappai$, while the remainder couples back into the central waveguide and is collected in reflection by the lensed optical fiber. For this particular cavity, we observe a total optical energy decay rate of $\kappa/2\pi=529$~MHz, an external coupling rate $\kappae/2\pi=153$~MHz, and an intrinsic decay rate $\kappai/2\pi=376$~MHz.  From the normalized reflection signal level ($R$), the fraction of optical power reflected by the OMC cavity, collected by the lensed fiber, and detected on a photodetector is estimated to be $\eta_{\text{cpl}}=34.7\%$, which after accounting for loss in external fiber-optic components, corresponds to a single pass fiber-to-waveguide coupling of $\eta_{\text{c}}=50\%$ (see App.~\ref{sec:appD} for more details).

The optical heterodyne detection scheme illustrated in Fig.~\ref{fig:setup}d is used to measure the motion of the localized 'breathing' mode at frequency $\omegamO/2\pi=3.6$~GHz, which is coupled to the fundamental optical resonance of the nanobeam. A high-power ($\sim$0.7~mW) local oscillator (LO) sets the gain of the heterodyne receiver, and a low-power ($2$~nW - $20$~$\mu$W) optical signal beam is used to probe the OMC cavity. In order to selectively detect either the upper or lower motional sidebands generated on the optical signal beam, the LO frequency ($\omegaLO$) is shifted relative to that of the signal beam ($\omegaS$) using a combination of a phase modulator and tunable optical filter. For the measurements presented here, $\omegaLO$ is adjusted such that the mechanical modulation beat frequency $\Omega = \omegaLO-(\omegaS \pm \omegamO) \approx 2\pi\times 50\text{~MHz}$, placing a single motional sideband within the bandwidth ($100$~MHz) of the balanced photodetectors of the heterodyne receiver. In the case of a signal beam red-detuned from cavity resonance (frequency $\omegacO$) by $\Delta \equiv \omegacO-\omegaS = \omegamO$, the resultant noise power spectral density (NPSD) as transduced on a spectrum analyzer yields a Lorentzian component proportional to $S_{\text{bb}}(\omega) = \nbar \gamma / ((\omega-\Omega)^2+(\gamma/2)^2)$~\cite{Safavi-Naeini2013a}. Here $\nbar$ is the phonon occupancy of the mechanical mode, and the total mechanical damping rate is given by $\gamma = \gammaiO + \gammaOMO$, where $\gammaiO$ is the intrinsic damping rate of the mechanical resonator and $\gammaOMO = 4 \gzeroO^2 \ncavO/\kappa$ is the optomechanically induced damping rate produced by an intracavity photon number $\ncavO$. For a blue-detuned probe ($\Delta = -\omegamO$), the NPSD Lorentzian is proportional to $\nbar+1$ and $\gamma = \gammaiO - \gammaOMO$.

Mechanical spectroscopy is first performed at a fridge temperature of $\Tf=4$~K in order to calibrate the optomechanical transduction. The coupling rate $\gzeroO$ is determined by observing the dependence of the mechanical linewidth on $\ncavO$ for both red ($\Delta=\omegamO$) and blue ($\Delta=-\omegamO$) laser-cavity detunings, as shown in Fig.~\ref{fig:4K}a.  Above a threshold value, $\ncavO > n_\text{thr} \approx 1.5$, optical amplification and self-oscillation of the mechanical resonator occurs for blue detuning.  Below this value, the optomechanical damping $\gammaOMO$ can be found from the difference between the red and blue detuned linewidths.  A linear fit of the derived $\gammaOMO$ versus $\ncavO$ yields a coupling rate of $\gzeroO/2\pi = 735$~kHz.   The mechanical mode occupancy versus $\ncavO$, plotted in Figure~\ref{fig:4K}b, is calibrated from the area under the Lorentzian part of the measured NPSD (see Fig.~\ref{fig:4K}c) using this value of $\gzeroO$, along with calibration of the total system efficiency, photodetector gain, and LO power.  At high $\ncavO$ significant cooling of the mechanical mode is measured for increasing $\ncavO$, whereas at low $\ncavO$ the calibrated occupancy saturates to a constant value, indicating thermalization to the fridge temperature. For these low probe powers, the calibrated mode occupancy is larger than the $4$~K fridge occupancy ($\nfridge$) by a factor of $\beta = 1.3$, which represents an unknown systematic error in our calibration procedure. This correction factor is applied to all subsequent mode occupancies presented below. 

%Figure~\ref{fig:4K}c shows a set of representative plots of the NPSD for a red-detuned probe, converted into units of displacement ($S_{\text{xx}} = x_{\text{zpf}}^2 S_{\text{bb}}$, where $x_{\text{zpf}}=4.1$~fm is the zero-point amplitude of the mechanical resonator). 

%At low $\ncavO$ ($\gammaOMO < \gamma$) the primary effect of increasing the probe power is to lower the noise floor while leaving the mechanical Lorentzian relatively unchanged, increasing the signal-to-noise ratio (SNR). As $\ncavO$ continues to increase, the optomechanical damping becomes significant, decreasing the total amount of noise power in the Lorentzian as well as increasing the bandwidth over which this noise power is distributed.

\begin{figure}[btp]
\begin{center}
\includegraphics[width=\columnwidth]{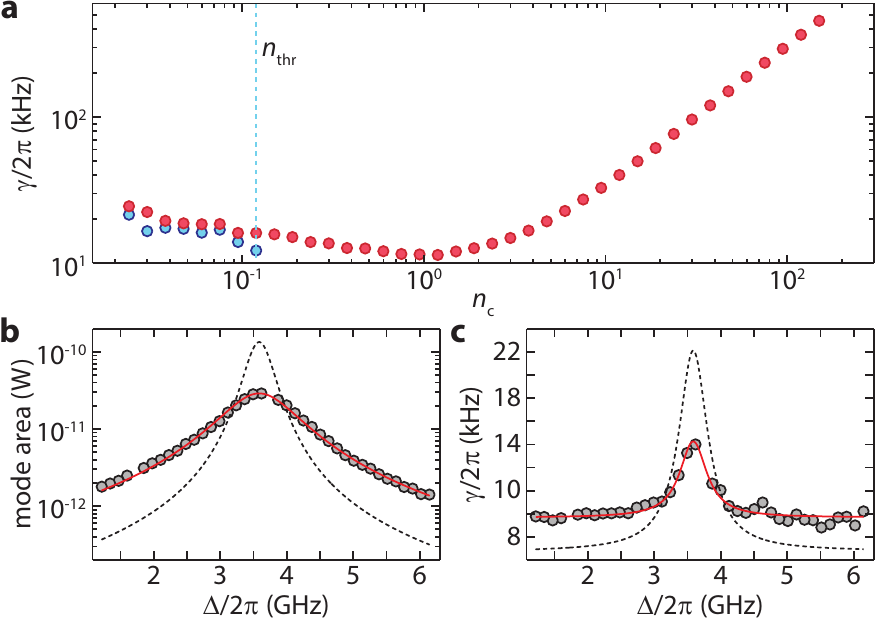}
\caption{\textbf{Linewidth and detuning sweeps at $\Tf$=185~mK} \textbf{a}, Mechanical linewidth vs.\ intracavity photon number for red- (red circles, $\Delta = \omegamO$) and blue-detuned (blue circles, $\Delta = -\omegamO$) probes at $\Tf=185$~mK. The blue dashed line indicates the threshold photon number where the onset of phonon lasing occurs. \textbf{b}, Measured (circles) mechanical mode area versus optical detuning, $\Delta$, for $\ncavO\sim 2$. The red curve shows a best fit to the data with $C=3.9$.  The black dashed curve shows the expected signal in the absence of optomechanical backaction cooling ($C \rightarrow 0$), consistent with assuming $\gammaiO$ is equal to the measured time-averaged linewidth. \textbf{c}, Measured (circles) mechanical linewidth versus $\Delta$ for $\ncavO \sim 2$. The red curve is a fit with $C$ constrained ($=3.9$), but assuming a Voigt lineshape with additional frequency jitter term.  The dashed black curve is the best fit for $C$ constrained ($=3.9$) but with no additional frequency jitter term.} \label{fig:185mK_detuning}
\end{center}
\end{figure}

As the fridge temperature is lowered into the sub-kelvin range, a very different dependence of measured linewidth and mode occupancy on optical probe power is observed.  In particular, the measured mechanical linewidth versus $\ncavO$, shown in Fig.~\ref{fig:185mK_detuning}a for $\Tf = 185$~mK, increases with decreasing probe power below an apparent minimum at $\ncavO \sim 1$.  The measured linewidth at low power is also too large to explain the observed threshold of self-oscillation, $n_{\text{thr}} \approx 0.1$. These inconsistencies indicate that the linewidth associated with the true energy decay rate of the mechanics is likely obscured in the time-averaged spectrum of Fig.~\ref{fig:185mK_detuning}a due to frequency jitter~\cite{Yang2011}.  Due to the long averaging times (minutes to hours) required at low optical probe power, direct observation of the frequency jitter in the 'breathing' mode is not possible.  Indirect confirmation and quantification of the frequency jitter, however, is possible by studying the detuning ($\Delta$) dependence of the mode occupancy and linewidth for a fixed $\ncavO \sim 2$.  Such a measurement keeps constant any effects such as optical heating or frequency jitter that might depend on the intra-cavity photon number.  From a fit to the integrated NPSD area shown in Fig.~\ref{fig:185mK_detuning}b (red curve), we extract a cooperativity at detuning $\omegamO$ of $C = \gammaOMO(\Delta=\omegamO)/\gammaiO = 3.9$.  This value of $C$ is then used as a constraint in fitting $\gamma(\Delta)$ assuming a Voigt lineshape (Fig.~\ref{fig:185mK_detuning}c), where the energy damping Lorentzian linewidth $\gamma_{\text{L}} = \gammaiO+\gammaOMO(\Delta)$ is fit with an additional random Gaussian frequency jitter term $\gamma_{\text{G}}$. The resulting fit (red curve) yields $\gammaiO/2\pi = 2.3$~kHz, $\gamma_{\text{G}}/2\pi=6.1$~kHz, and $\gzeroO/2\pi = 715$~kHz consistent with the $\Tf = 4$~K value.  In what follows we use additional on-resonance heating measurements to determine $\gammaiO$ over the entire range of $\ncavO$, including at the lowest optical probe powers where back action is weak.

%These fit values are both consistent with the measured threshold for self-oscillation and the $\Tf=4$~K value for $\gzeroO$. Note that this technique is not effective in determining $\gammaiO$ at the lowest optical probe powers where back-action is weak, and so in what follows we use additional on-resonance heating measurements to determine $\gammaiO$ over the entire range of $\ncavO$. 

%If the optomechanical cooling of the mechanics is neglected ($C \ll 1$), as is expected for an intrinsic mechanical damping rate equal to the measured time-averaged linewidth at $\ncavO=2$, the dependence of the integrated NPSD area on $\Delta$ (Fig.~\ref{fig:185mK_detuning}b) should follow the optical cavity spectrum with linewidth $\kappa$ (dashed black curve). However, f

\begin{figure*}[btp]
\begin{center}
\includegraphics[width=2\columnwidth]{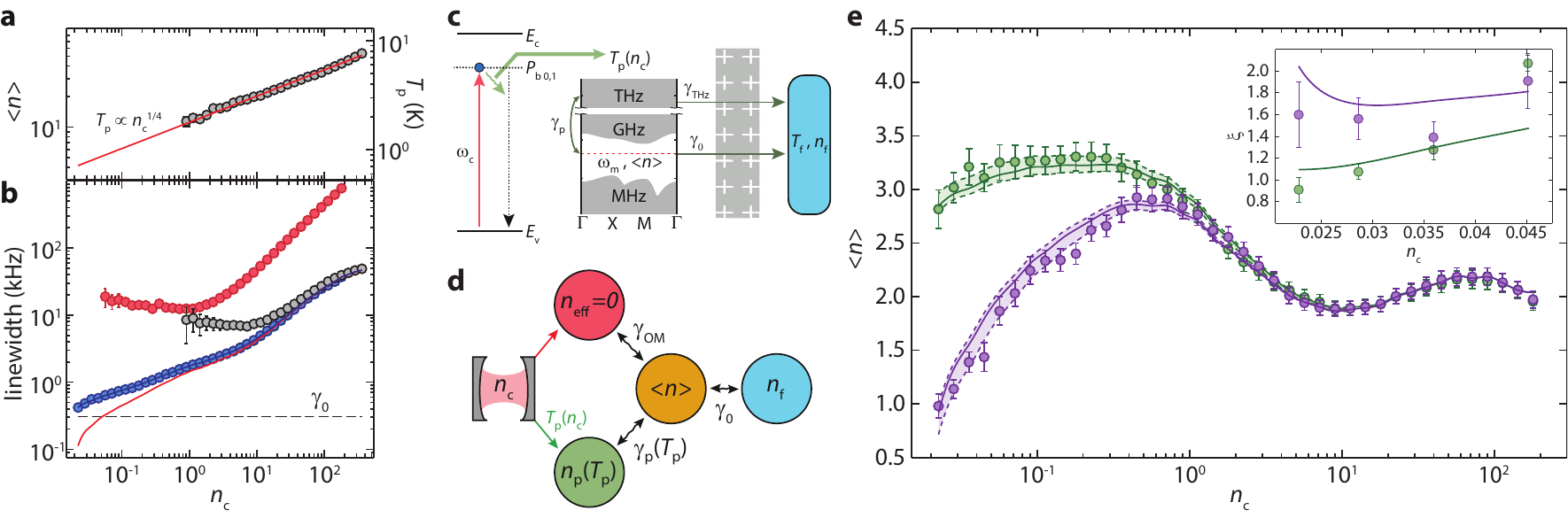}
\caption{\textbf{Optical Absorption Heating.}  \textbf{a}, Measured phonon occupancy versus intracavity photon number for a resonant probe at $\Tf=10$~mK. The red line shows a power law fit to the occupancy, $\nbathp$. The right axis shows the equivalent bath temperature, $\Tbathp$.  \textbf{b}, Measured mechanical linewidth versus $\ncavO$ for resonant (gray circles) and red-detuned (red circles) probes. The blue circles show the best fit values of $\gammaiO = \gammanotO + \gammapO$ to both $\Tf=10$ and $635$~mK data sets.  The best fit value for $\gammanotO$ and a smooth curve fit to the inferred $\gammapO$ are shown as a black dashed line and a solid red line, respectively.  \textbf{c}, Diagram illustrating the proposed model of optical-absorption-driven heating of the mechanics and \textbf{d}, schematic showing the various baths coupled to the localized mechanical mode.  See text for details.  \textbf{e}, Calibrated mode occupancy versus $\ncavO$ for $\Tf=10$~mK and $635$~mK.  A best fit to both temperature data sets using the proposed heating model are shown as solid curves, with shaded region representing the variation in the fit for $\gammanotO/2\pi = 306 \pm 28$~Hz.  The inset shows the measured asymmetry $\xi$ as a function of $\ncavO$, with the predicition of the best fit model shown as solid curves.  $\Tf=10$~mK data/fits are shown as purple circles/solid curves.  $\Tf=635$~mK data/fits are shown as green circles/solid curves.} \label{fig:resonant}
\end{center}
\end{figure*}

Heating of the mechanical mode by optical absorption becomes significant at sub-kelvin temperatures due to the sharp drop in thermal conductance with temperature~\cite{Holland1963}.  The source of optical absorption in our structures is most likely due to electronic defect states at the surface of Si~\cite{Stesmans1996,Borselli2007}, with phonon-assisted relaxation from these mid-gap states heating the mechanics. This heating mechanism is investigated at $\Tf=10$~mK, for which $\nfridge$ is negligible, by measuring $\nbar$ and $\gamma$ using an optically resonant probe ($\Delta = 0$; $\gammaOMO =0$).  We observe in Fig.~\ref{fig:resonant}a that $\nbar \propto \ncavO^{1/4}$.  This weak power law dependence is consistent with indirect coupling of the 'breathing' mode to an optically-generated bath with thermal conductance scaling as $G_\text{th} \sim T^3$.  High frequency phonons with wavelengths small relative to the dimensions of the cavity structure should have this conductance scaling~\cite{Chen2008b}, although their expected fast escape time ($\gamma_{\text{THz}}^{-1}\sim 1\mbox{--}10$~ns) suggests that they quickly come into thermal equilibrium. Given the slow rates of most bulk relaxation processes, such as anharmonic three-phonon mixing at low temperatures~\cite{Mingo2003}, this fast thermalization is likely due to a relatively large degree of diffusive and inelastic scattering at the surfaces of the patterned nanobeam~\cite{Hertzberg2013}.  

%Figure~\ref{fig:resonant}b shows the measured on-resonance $\gamma$ ($=\gammaiO$) versus $\ncavO$ (gray circles), which for linewidths larger than the spectral diffusion limit ($\ncavO \gtrsim 10$), also shows a strong depenence on intra-cavity photon number.

% In the low-probe-power regime explored in this work, the excitation of free-carriers due to two-photon absorption has been found to be negligible by previous stuides of Si OMCs~\cite{Chan2011}. 

% Calibrated mechanical occupation for a red detuned ($\Delta=\omegamO$) probe is plotted against $\ncavO$ in Fig.~\ref{fig:resonant}a for dilution refrigerator base temperatures $\Tf=10$~mK (purple) and $635$~mK (green). Both curves exhibit a series of heating and cooling trends, and in fact coincide for $\ncavO \geq 1$, indicating that at these powers the $\ncavO$-dependent heating dominates the mechanical occupation irrespective of the fridge temperature. 

%The coupling rate $\gammapO$ to this optically-generated thermal bath is also measured by probing on-resonance ($\Delta=0$).  As the local temperature increases, $\gamma$ is expected to increase due to the larger population of phonons which can inelastically scatter with the mechanical resonator mode. This is observed in the on-resonance dependence of $\gamma$ even in the absence of optomechanical damping (Fig.~\ref{fig:resonant}b). 

Based on these on-resonance observations, a proposed microscopic model for the optical absorption heating and damping is illustrated in Fig.~\ref{fig:resonant}c.  Here the long-lived 'breathing' mode is weakly coupled ($\gammanotO$) through the phononic shield to the exterior fridge environment, and is locally coupled via phonon-phonon scattering ($\gammapO$)~\cite{Zyryanov1966} to the optically-generated high frequency phonons within the acoustic cavity.  A phenomenological model based upon this microscopic picture is shown schematically in Fig.~\ref{fig:resonant}d.  We parameterize the coupling of the mechanical resonator to the separate thermal baths by decomposing the mechanical damping rate into $\gamma_{\text{L}} = \gammanotO + \gammapO + \gammaOMO$, where the fridge bath (occupancy $\nfridge$) couples at rate $\gammanotO$, the optical-absorption-induced bath (temperature $\Tbathp$, occupancy $\nbathp$ at $\omegamO$) couples at rate $\gammapO$, and the intra-cavity laser field (effective zero-temperature bath) couples at rate $\gammaOMO$. The resulting average mechanical mode occupation is then given by $\nbar(\ncavO) = (\gammanotO \nfridge + \gammapO(\Tbathp) \nbathp (\Tbathp))/(\gammanotO + \gammapO (\Tbathp) + \gammaOMO (\ncavO))$, where $\Tbathp(\ncavO)$.

The calibrated mechanical mode occupation for a red detuned ($\Delta=\omegamO$) probe is plotted against $\ncavO$ in Fig.~\ref{fig:resonant}e for $\Tf=10$~mK (purple) and $635$~mK (green).  Both curves exhibit a series of heating and cooling trends, and in fact coincide for $\ncavO \gtrsim 1$.  At the lowest optical probe powers ($\ncavO=0.021$) and lowest fridge temperature ($\Tf=10$~mK), the calibrated phonon occupancy reaches a minimum $\nbar = 0.98 \pm 0.11$, corresponding to $T \approx 270$~mK.    The complex behavior of these two cooling curves can be understood by comparing to the proposed phenomenological model.  For this model $\nfridge$ is taken to correspond to the measured $\Tf$, $\nbathp(\ncavO)$ is ascertained by extrapolating the on-resonance measurement of $\nbar$ (Fig.~\ref{fig:resonant}a), and $\gammaOMO(\ncavO)$ is found from the fit value $\gzeroO/2\pi=735$~kHz to the high power region of the red detuned mechanical linewidth (red circles in Fig.~\ref{fig:resonant}b).  Assuming a common $\gammapO(\ncavO)$ and $\gammanotO$, the resulting $\gammaiO(\ncavO)$ curve that best fits the measured $\nbar$ data for both $\Tf=10$~mK and $635$~mK fridge temperatures is plotted in Fig.~\ref{fig:resonant}b (blue circles).  Also shown in Fig.~\ref{fig:resonant}b are the best fit value of the coupling to the fridge bath $\gammanotO/2\pi = 306 \pm 28$~Hz (dashed black horizontal curve) and a smooth spline curve fit to the inferred values of $\gammapO(\ncavO)$ (red solid curve).  A plot of the best fit model is shown alongside the measured $\nbar$ cooling curves in Fig.~\ref{fig:resonant}e.  In addition to the good agreement of the model for both fridge temperatures, the inferred intrinsic energy damping rate at $\ncavO \sim 2$ is consistent with that from the above analysis of the detuning dependence of the transduced mechanical signal and the measured linewidth.  The estimate of the threshold for self-oscillation based upon the inferred $\gammaiO$ is also consistent with the measured threshold in Fig.~\ref{fig:185mK_detuning}a.  At the lowest probe powers ($\ncavO =0.021$), the energy damping mechanical $Q$-factor reaches an impressively high value of $\QmO = 9 \times 10^{6}$.

Alongside the calibrated mode occupancy $\nbar$, we have also measured the sideband asymmetry, $\xi$, shown in the inset to Fig.~\ref{fig:resonant}e.  The sideband asymmetry is defined as $\xi = I_{-}/I_{+} - 1$~\cite{Safavi-Naeini2012}, where $I_{\pm}$ is the area under the Lorentzian part of the NPSD for an optical probe with detuning $\Delta = \pm \omegamO$.  The asymmetry is sensitive to both the absolute mode occupancy and to the sum of $\gammanotO$ and $\gammapO$ through the cooperativity $C = \gammaOMO/(\gammanotO+\gammapO)$.  Good correspondence can also be seen between the best fit model (solid curves) and the measured $\xi$ (circles).      

Although significant work remains to determine the exact microscopic details of the optical absorption heating, thermalization properties, and frequency jitter observed in the measurements of the quasi-1D OMC cavity studied here, there are nonetheless several interesting points that can already be noted.  First, from the measured time-averaged mechanical linewidth and optically-induced bath temperature ($\Tbathp$) in Fig.~\ref{fig:resonant} we find that the frequency noise of the mechanical resonator drops with increasing temperature as $\Tbathp^{-0.9}$ (see App.~\ref{sec:appG} for details).  Such a frequency noise behavior is similar to that found for two level systems (TLS) coupled to superconducting microwave resonators~\cite{GaoPhD}.  The presence of TLS, common in amorphous materials, can be expected in the devices of this work due to native oxide formation at the Si surfaces.  Secondly, despite the very high optical quality of the Si devices, optical absorption still plays a significant role in heating and damping of the local 'breathing' mode at sub-kelvin temperatures.  Here, the phononic shield provides excellent mechanical isolation of the 'breathing' mode, yet still provides good mechanical coupling to the fridge bath for heat carrying phonons above the acoustic bandgap.  Thirdly, although lower phonon occupancies could have been measured using thinner phononic shields, effectively increasing the coupling rate $\gammanotO$ to the fridge bath at $\Tf$, this would come with a commensurate reduction in cooperativity $C=\gammaOMO/\gammaiO$.  Ultimately, coherent quantum interactions between the optical cavity field and the mechanical resonator require $C > 1$ and $\nbar < 1$.  The devices of this work closely approach this limit, however, quasi-2D Si OMC devices such as those recently realized in Ref.~\cite{Safavi-Naeini2013} should have orders of magnitude larger thermal conductance, enabling coherent coupling between light and mechanics in the quantum regime as envisioned in recent proposals~\cite{Schmidt2012,Tomadin2012,Ludwig2013,Schmidt2013}.

\begin{acknowledgements}
The authors would like to thank Michael Roukes, Ron Lifshitz, and Michael Cross for helpful discussions regarding the proposed thermal model, as well as Jasper Chan, Witlef Wiezcorek, and Jason Hoelscher-Obermaier for support in the early stages of the experiment.  This work was supported by the DARPA ORCHID and MESO programs, the Institute for Quantum Information and Matter, an NSF Physics Frontiers Center with support of the Gordon and Betty Moore Foundation, and the Kavli Nanoscience Institute at Caltech.  ASN acknowledges support from NSERC. SG was supported by a Marie Curie International Outgoing Fellowship within the $7^{\textrm{th}}$ European Community Framework Programme.
\end{acknowledgements}

\bibliographystyle{apsrev4-1}
%\bibliography{Mirror_v2}


\appendix

\section{Optical and mechanical design}
\label{sec:appA}

\begin{figure*}[btp]
\begin{center}
\includegraphics[width=2\columnwidth]{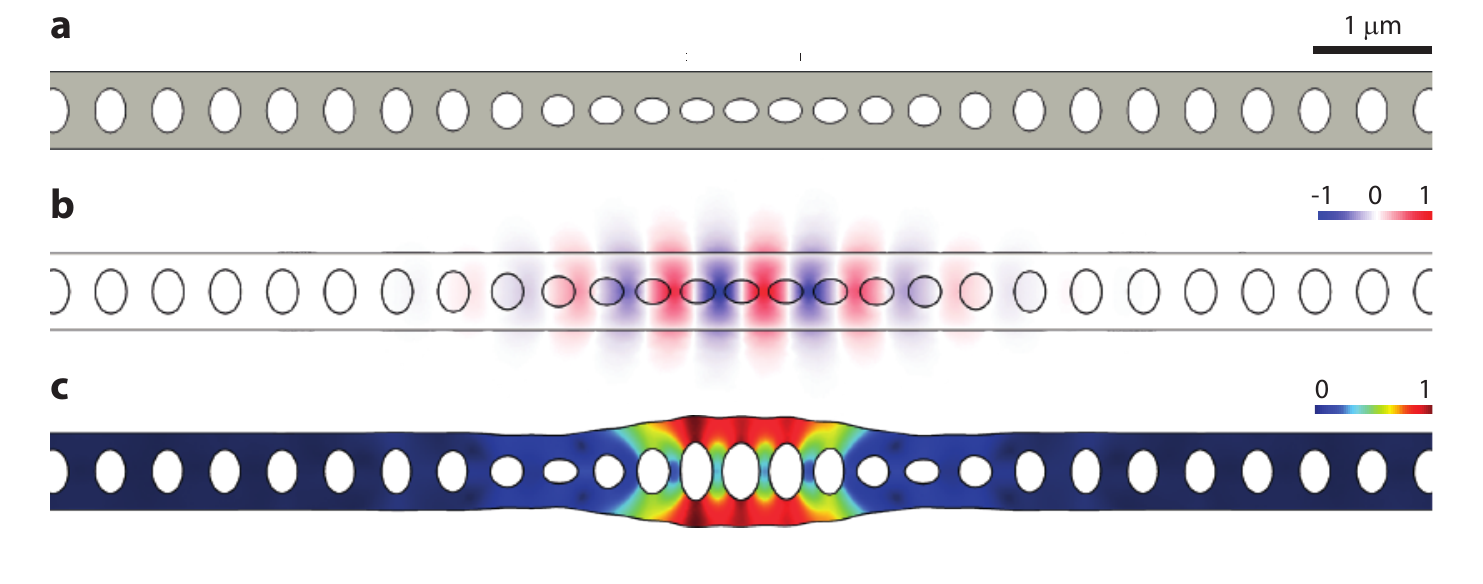}
\caption{\textbf{Optical and Mechanical FEM Simulations.} \textbf{a}, Silicon OMC geometry.  \textbf{b}, Optical mode FEM simulation, showing the electric field $E_y$ component (polarization in the plane of the page and transverse to the long axis of the nanobeam). \textbf{c}, FEM simulation of the localized 'breathing' mode at frequency $\omegamO/2\pi = 3.6$~GHz.  Here the displacement mode profile is shown as a distortion of the structure and as a color-coded local displacement amplitude.} \label{Sfig:sims}
\end{center} 
\end{figure*}

The geometry of the optomechanical crystal (OMC) studied in this work is numerically optimized for optical and mechanical quality, as well as optomechanical coupling, via finite-element method (FEM) simulation in COMSOL Multiphysics~\cite{COMSOL}. In Fig.~\ref{Sfig:sims}a, a top view of the OMC shows the center defect in the $600$~nm wide by $220$~nm thick Si nanobeam. The larger holes on the ends of the nanobeam support simultaneous bandgaps for $1550$~nm band light and $3-4$~GHz acoustic waves, while the smaller holes in the center of the nanobeam perturb the bandgaps such that optical and mechanical modes are co-localized~\cite{Chan2012}. The fundamental optical mode has a nominal wavelength of $1535$~nm (Fig.~\ref{Sfig:sims}b), and the acoustic breathing mode has a nominal resonance frequency of $3.85$~GHz (Fig.~\ref{Sfig:sims}c). Through simulation of the dependence of the effective index of refraction of the structure on the mechanical mode through moving-boundary and photo-elastic effects, we determine the nominal optomechanical vacuum coupling rate $\gzeroO$ to be $870$~kHz. This constant represents the shift in optical resonance frequency due to zero-point fluctuations of the mechanical resonator. In fabrication, arrays of the nominal design in Fig.~\ref{Sfig:sims} are scaled by $\pm~2\%$ to account for geometrical imperfections, leading to a range of realized optical and mechanical resonance frequencies. For the particular device studied here, the optical wavelength is $1545$~nm, the mechanical frequency is $3.6$~GHz, and $\gzeroO=735$~kHz, as discussed in the main text.

\section{Fabrication}
\label{sec:appB}

The devices are fabricated from a silicon-on-insulator (SOI) wafer (SOITEC, 220 nm device layer, 3 $\mu$m buried oxide) using electron beam lithography followed by reactive ion etching (RIE/ICP). The Si device layer is then masked using ProTEK PSB photoresist to define a mesa region of the chip to which a tapered lensed fiber can access. Outside of the protected mesa region, the buried oxide is removed with a plasma etch and a trench is formed in the underlying silicon substrate using tetramethylammonium hydroxide (TMAH). The devices are then released in hydrofluoric acid (49\% aqeuous HF solution) and cleaned in a piranha solution (3-to-1 H$_2$SO$_4$:H$_2$O$_2$) before a final hydrogen termination in diluted HF.

\section{Experimental setup}
\label{sec:appC}

\begin{figure*}[btp]
\begin{center}
\includegraphics[width=2\columnwidth]{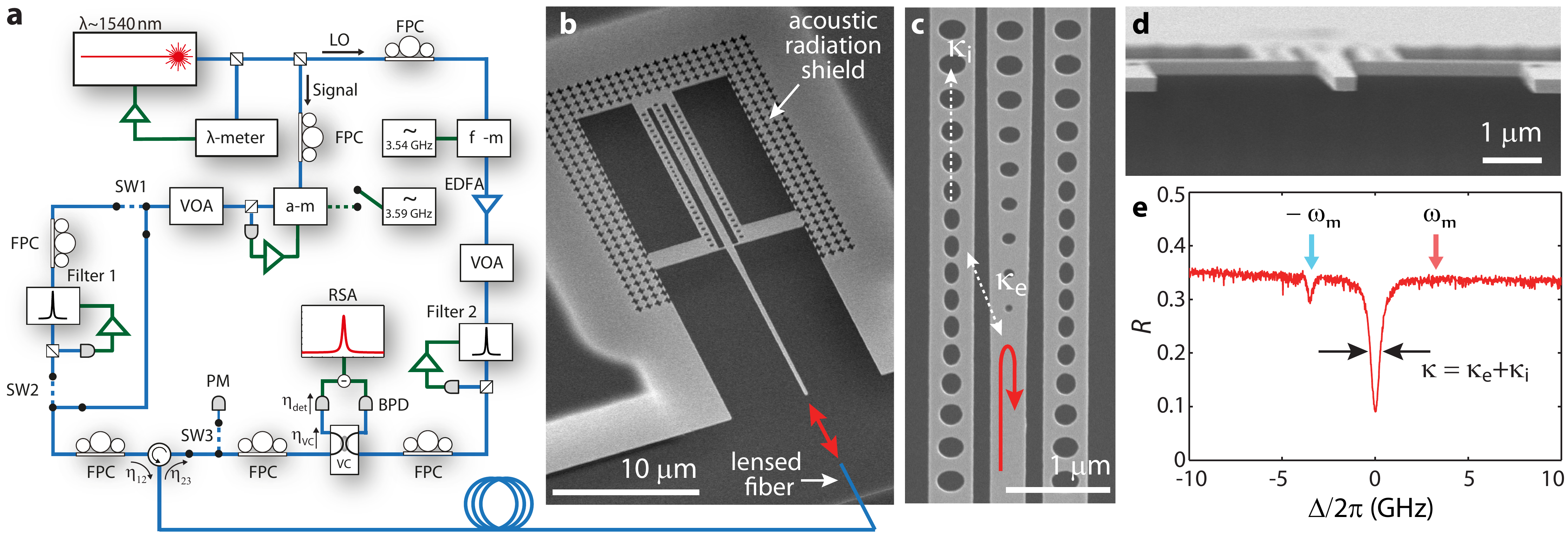}
\caption{\textbf{Experimental setup.}\textbf{a}, Fiber-based optical heterodyne setup. $\lambda$-meter:\ wavemeter, FPC: fiber polarization controller, a-m: electro-optic amplitude modulator, $\varphi$-m:\ electro-optic phase modulator, VOA:\ variable optical attenuator, SW: optical switch, EDFA: erbium-doped fiber amplifier, PM: optical power meter, VC:\ variable coupler, BPD: balanced photodiode pair, RSA:\ real-time spectrum analyzer. \textbf{b},  SEM image of the OMC device, with acoustic radiation shield and end-fire fiber coupling indicated. \textbf{c}, Zoom in SEM image showing details of the waveguide--cavity coupling region, indicating the extrinsic cavity-waveguide coupling rate $\kappae$ and the intrinsic loss rate $\kappai$. \textbf{d}, Edge view SEM image of the central waveguide tip. \textbf{e},  Normalized optical cavity reflection spectrum ($R$). Detunings from resonance of $\Delta = \pm\omegamO/2\pi=\pm 3.6$~GHz are denoted by the red and blue arrows, respectively.} \label{Sfig:setup}
\end{center} 
\end{figure*}

The full experimental setup for heterodyne spectroscopy and mechanical thermometry is shown in Fig.~\ref{Sfig:setup}a. A fiber-coupled, wavelength-tunable external cavity diode laser is used as the light source, and a small percentage of the laser output is sent to a wavemeter ($\lambda$-meter) for frequency stabilization. The remaining laser power is split into a high-power ($\approx$0.7~mW) local oscillator (LO) path and a low-power ($\approx$20~$\mu$W) signal path. The signal beam is sent through an electro-optic modulator (a-m) to stabilize the signal intensity and a variable optical attenuator to allow control of the probe power sent to the cavity. The signal is sent into an optical circulator which directs the signal beam to the dilution refrigerator in which the fiber terminates with a lensed tip for end-fire coupling to the device. The cavity reflection then circulates to a variable coupler (VC) and mixes with the LO before being detected on a pair of balanced photodiodes (BPD). The difference photocurrent is then amplified and its noise power spectral density (NPSD) is measured on a real-time spectrum analyzer (RSA). The LO is first sent through an electro-optic phase modulator ($\varphi$-m) which produces optical sidebands at $\pm$($\omega_{\text{m}}/2\pi - 50$~MHz) for the purpose of mixing the mechanically induced signal modulation down to within the $100$~MHz bandwidth of the BPD circuit. An erbium-doped fiber amplifier (EDFA) and a variable optical attenuator (VOA) are used to set the power of the LO sidebands, and the appropriate sideband is selected by a high-finesse tunable Fabry-P\'{e}rot filter before recombining it with the signal.

\begin{figure*}[btp]
\begin{center}
\includegraphics[width=2\columnwidth]{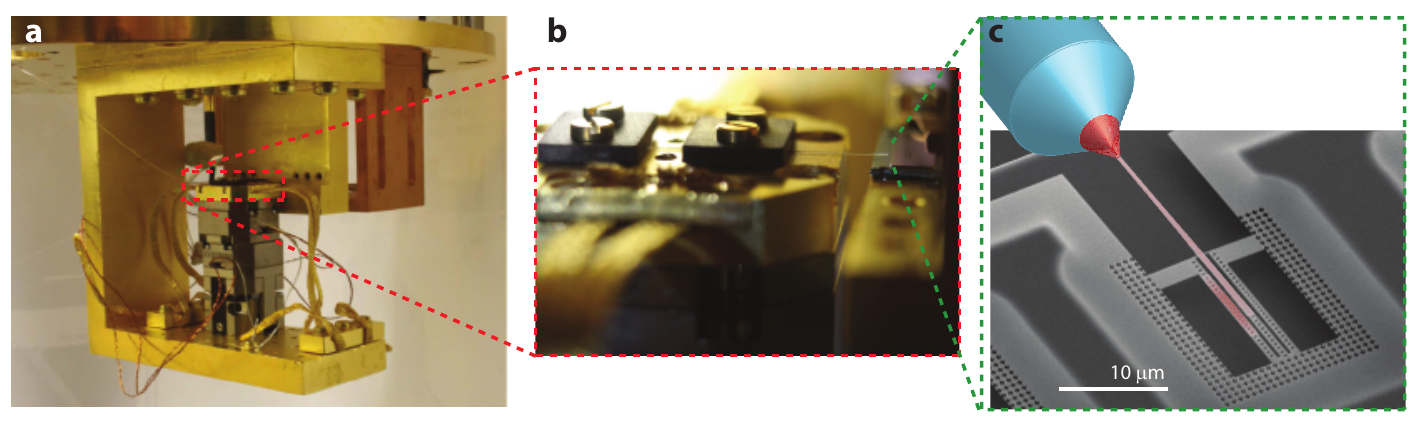}
\caption{\textbf{Sample mounting and fiber coupling in the dilution refrigerator.} \textbf{a}, Photo of the attocube nanopositioning stages mounted to the bottom of the dilution refrigerator mixing plate. \textbf{b}, Photo of the lensed fiber tip mounted on the attocube positioners, illustrating alignment to a Si test chip. \textbf{c}, Diagram showing the lensed fiber (not-to-scale) coupling to a silicon OMC shown in the SEM image, with the superimposed optical mode FEM simulation of optical intensity.} \label{Sfig:photos}
\end{center} 
\end{figure*}

Fig.~\ref{Sfig:setup}b shows a scanning electron micrograph (SEM) of a typical device, consisting of a central Si waveguide side-coupled to two nanobeam OMCs, all of which is surrounded by an acoustic radiation shield phononic crystal. Light coupled into the central waveguide either reflects from a photonic crystal mirror or, when resonant, evanescently couples to one of two side-coupled OMCs~\cite{Chan2011,Groeblacher2013a}, where the coupling rate $\kappa_{\text{e}}$ is controlled by the design of the gap size between the waveguide and the cavity (Fig.~\ref{Sfig:setup}c). Some cavity light decays through parasitic scattering at a rate $\kappa_{\text{i}}$, while the remainder couples back into the central waveguide and is collected from the central waveguide tip (Fig.~\ref{Sfig:setup}d) by the lensed fiber to be guided to the photodetector in Fig.~\ref{Sfig:setup}a. By tuning the laser over the cavity mode, we observe in the reflected signal on the output of the dilution refrigerator (Fig.~\ref{Sfig:setup}e) a fiber collection efficiency of $\eta_{\text{cpl}}=34.7\%$ and an optical linewidth of $\kappa/2\pi=529$~MHz, composed of $\kappa_{\text{e}}/2\pi=153$~MHz and $\kappa_{\text{i}}/2\pi=376$~MHz.

\section{Fiber coupling in the dilution refrigerator}
\label{sec:appD}

The microchip sample is mounted to the mixing chamber of the dilution refrigerator, and we utilize an end-fire coupling scheme to probe individual devices with an anti-reflection-coated tapered lensed fiber. The lensed fiber tip is clamped down on a position encoded piezo xyz-stage inside the dilution refrigerator (Fig.~\ref{Sfig:photos}a), which allows nanopositioning with respect to the sample. When mounting the fiber and sample, the fiber is only roughly aligned to within a few millimeters (Fig.~\ref{Sfig:photos}b). After cooling the experiment from room temperature to $4$~K, we monitor the reflected optical power on a slow photodetector as we carefully lower the fiber tip to match the height of the device layer. The distinct reflection of the device layer allows us to iteratively adjust the fiber position and optimize the coupling to each device (Fig.~\ref{Sfig:photos}c).

The fiber-tip launches the light to free-space and focuses it to a beam waist of 2.5~$\mu$m at a focal distance of 14~$\mu$m. We position the fiber such that the beam waist aligns to a silicon waveguide tip (Fig.~\ref{Sfig:setup}b,d) that matches the convergence of the optical field, which becomes a guided mode in the waveguide ~\cite{Mitomi1994,Almeida2003,Cohen2013}.

The design of the tapered waveguide coupler is similar to that presented in Ref.~\cite{Cohen2013}, where the tip of the waveguide is mode matched to an input Gaussian field of the appropriate width and adiabatically tapered up to the full width of the photonic crystal mirror section. The major distinction is the use of end-fire coupling utilizing a lensed fiber rather than butt-coupling with a cleaved single-mode fiber. In contrast to the high-stress Si$_3$N$_4$ utilized in Ref.~\cite{Cohen2013} a Si waveguide will begin to sag if the waveguide is made too long, even in the presence of a supporting tether, which will lead to misalignment of the waveguide tip and bending losses that inhibit the coupling efficiency. As such, it is necessary to substantially reduce the length of the waveguide taper, which necessitates a smaller overall change in waveguide width to maintain the adiabaticity of the taper. Consequently we are forced to use a larger waveguide width at the tip and a correspondingly smaller mode field diameter such as can be obtained with a lensed fiber.

\begin{figure*}[btp]
\begin{center}
\includegraphics[width=2\columnwidth]{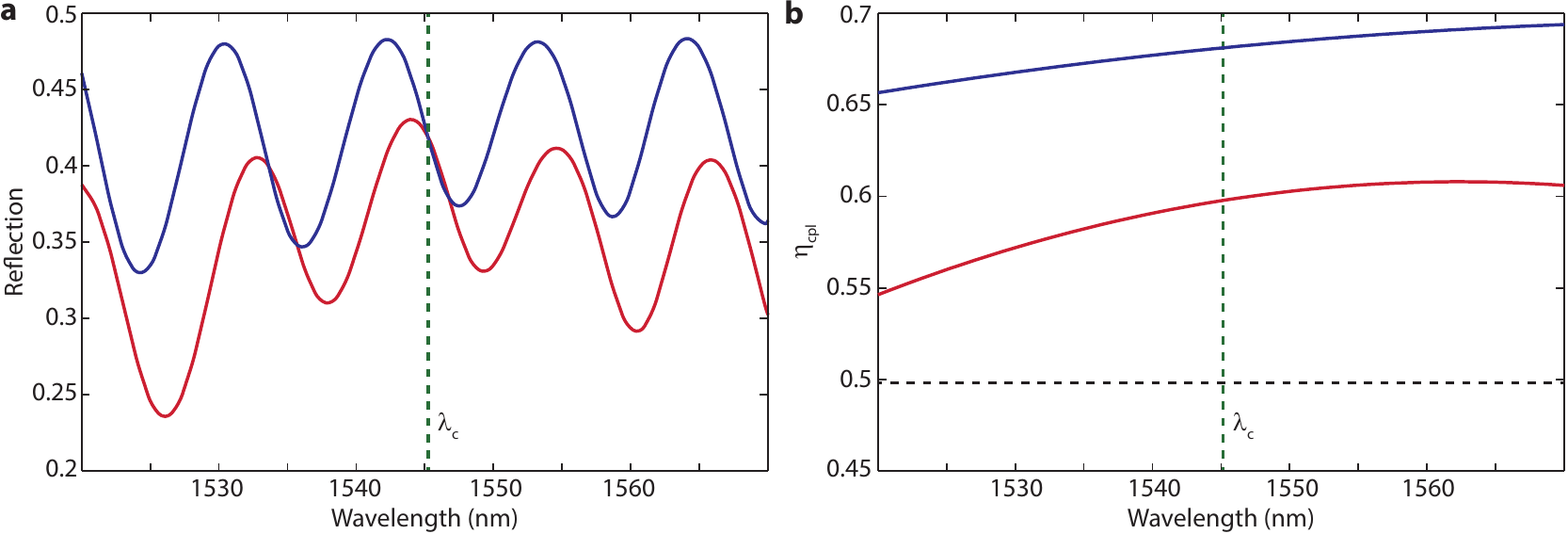}
\caption{\textbf{Coupling efficiency.} \textbf{a}, The simulated reflection of the tapered waveguide coupler as a function of wavelength, for an ideal (blue) coupler and a coupler with the actual dimensions of the measured device (red). The resonant wavelength of the measured optical cavity is indicated by a dashed green line. \textbf{b}, The theoretical single pass reflection efficiency, determined from the simulated curves in \textbf{a}. The black dashed line indicates the measured single pass efficiency of the device used in this work.} \label{fig:coupler}
\end{center} 
\end{figure*}

The broadband reflection spectrum of the coupler, calculated using finite-difference-time-domain simulation~\cite{Lumerical}, is shown for an ideal coupler in Fig.~\ref{fig:coupler}a. The presence of fringes in the reflection spectrum are consistent with the estimated weak reflectivity of the waveguide--air interface ($R \approx 0.5$\%), which forms a low-finesse Fabry-P\'{e}rot cavity with the high reflectivity photonic crystal mirror at the other end of the waveguide. From the fringe visibility we can back out the expected single-pass coupling efficiency $\eta_\text{cpl}$ shown in Fig.~\ref{fig:coupler}b~\cite{Cohen2013}.

The calculated reflection spectrum and single-pass efficiency for the device used in this work (using dimensions determined from SEM images) is also shown in Fig.~\ref{fig:coupler}a,b and is expected to be $\eta_\text{cpl} \approx 60$\% at the cavity resonance. The actual single-pass efficiency of the measured device is found to be $\eta_{cpl} \approx 50$\%. The difference from simulations is attributed to the difficulty of measuring the exact dimensions of the sample using the SEM, as a small difference of even 10 nm in the width of the waveguide tip can have a significant effect on the mode matching and thus on the overall coupling efficiency. 

\section{Calibration of the optical transduction of mechanical motion}
\label{sec:appE}

To perform accurate thermometry of the mechanical mode, it is necessary to calibrate the detection efficiency of the setup. We first measure the efficiency of transmission in the circulator from port 1 to port 2 ($\eta_{\text{12}} = 88 \%$), and from port 2 to port 3 ($\eta_{\text{23}} = 84 \%$). These values are measured once when the optical components are connected and do not change. These calibrations are used to determine the reflection efficiency of the device and the overall detection efficiency of the heterodyne setup. 

To measure device efficiency, the laser is tuned off-resonance from the optical mode (where the device should act as a near-perfect mirror) and a continuous-wave signal of input power $P_{\text{in}}$ is sent into port 1 of the circulator, leading to a power $\eta_{\text{12}} P_{\text{in}}$ exiting port 2 of the circulator. The optical losses incurred in the path from port 2 to the device-under-test are accumulated into an efficiency factor $\eta_{\text{cpl}}$, which includes signal loss in the fiber path through the fridge, mode-mismatch between the lensed-fiber to tapered-waveguide tip, and mode-scattering from the waveguide tip to the photonic-crystal mirror (Fig.~\ref{Sfig:photos}c). These losses are incurred twice in reflection back to the circulator, so a power of $\eta_{\text{cpl}}^2 \eta_{\text{12}} P_{\text{in}}$ propagates into port 2 and $\eta_{\text{23}} \eta_{\text{cpl}}^2 \eta_{\text{12}} P_{\text{in}}$ propagates out of port 3 of the circulator. An optical switch (SW3) is used to send this signal to a power meter (PM), and thus the coupling efficiency is determined as

\begin{equation}
\eta_{\text{cpl}} = \sqrt \frac{P_{\text{PM}}} {\eta_{\text{23}}  \eta_{\text{12}} P_{\text{in}}} = 34 \% .
\end{equation}

To calibrate the overall detection efficiency, we must also determine the efficiency of the heterodyne detector itself, which includes the intrinsic quantum efficiency of the BPD, the alignment of the polarization between the LO and the signal, and the degree to which the LO power overcomes the electronic noise of the detector. This is accomplished by using the amplitude modulator to create optical sidebands detuned from the signal by the mechanical frequency while the laser is tuned off-resonance from the optical mode. The optical switches SW1 and SW2 are used to route the signal through a tunable filter to select a single sideband which is sent through the device and onto the BPD. The power $P_{\text{cal}}$ in this sideband is directly measured on the PM at SW3, and the photocurrent NPSD ($S_{\text{II}}(\omega)$) as transduced on the RSA is given by 

\begin{equation}
S_{\text{II}}(\omega) = S_{\text{dark}} + \frac{G_{\text{e}}^2 }{R_{\text{L}}}S_{\text{SN}}^2 \left( 1 + \frac{\eta_{\text{VC}}\eta_{\text{det}}S_{\text{cal}}(\omega)}{\hbar\omega_{\text{o}}}  \right) ,
\end{equation}

\noindent where $S_{\text{dark}}(\omega)$ is the electronic NPSD of the detector, $S_{\text{SN}} = \sqrt{2\hbar\omega_{\text{o}}P_{\text{LO}}}$ is the optical shot-noise NPSD arising from $P_{\text{LO}}$ of LO optical power at optical frequency  $\omega_{\text{o}}$, which lies an order of magnitude above the electronic noise, and $S_{\text{cal}}$ is the NPSD of the signal, where $\int_{-\infty}^{\infty}  S_{\text{cal}}(\omega) \frac{d\omega}{2\pi} = P_{\text{cal}}$. The gain factor $G_{\text{e}}$ represents the conversion from optical power to voltage while $R_{\text{L}}$ is the input impedance of the RSA. The total noise floor $S_{\text{noise}} = \frac{G_{\text{e}}^2 }{R_{\text{L}}}S_{\text{SN}}^2+ S_{\text{dark}}$ is measured with the signal beam blocked, while $S_{\text{dark}}$ is measured independently with both signal and LO beams blocked. When referenced back to the PM, the calibration tone (with NPSD $S_{\text{cal}}(\omega)$) picks up losses in the VC and BPD, as parametrized in $\eta_{\text{VC}}$ and $\eta_{\text{det}}$, respectively. The efficiency of the heterodyne receiver is extracted as

\begin{equation}
\eta_{\text{VC}}\eta_{\text{det}}  = \frac{\hbar\omega_{\text{o}}}{P_{\text{cal}}} \int_{-\infty}^{\infty} \frac{S_{\text{II}}(\omega) - S_{\text{noise}}}{S_{\text{noise}}- S_{\text{dark}}} \frac{d\omega}{2\pi} = 56 \% .
\end{equation}

This, combined with the measured device coupling efficiency, yields the overall measurement efficiency $\eta$ used for calibrated mechanical thermometry as

\begin{equation}
\eta = \eta_{\text{cpl}} \eta_{\text{23}} \eta_{\text{VC}} \eta_{\text{det}} = 16 \% \text{.}
\end{equation}

\section{Heating and damping via three-phonon scattering processes}
\label{sec:appF}

Though a detailed microscopic calculation of the additional heating and damping due to optical absorption is beyond the scope of this work, some qualitative insight into the nature of the locally heated mechanical bath and its coupling to the mechanical mode of interest can be obtained from consideration of a simplified model of the phonon-phonon interactions. At the temperatures considered in this work ($T < 10$~ K) the mean free path of the thermal phonons is expected to be much larger than the wavelength of the mechanical mode of interest. Consequently, the damping should be described by the Landau-Rumer theory, where losses occur primarily due to three-phonon mixing with the local thermal environment due to anharmonicity in the Si lattice~\cite{Zyryanov1966,Srivastava1990}.

Consider first a toy model where two high-frequency modes, with frequencies $\omega_\text{1}$ and $\omega_\text{2}$ respectively, are coupled with the mode of interest at frequency $\omega_\text{m}$ ($\omega_\text{1}-\omega_\text{2} = \omega_\text{m}$). To first order in perturbation theory, the scattering rates into and out of the mechanical cavity mode due to the lowest order anharmonic interaction can be given by~\cite{Srivastava1990} $\Gamma_\text{+} = A (n_\text{m}+1)(n_\text{2}+1)n_\text{1}$ and $\Gamma_\text{-} = A (n_\text{1}+1) n_\text{m} n_\text{2}$, where $A$ is a constant which depends on the matrix element of the anharmonic potential, and $n_\text{1}$, $n_\text{2}$ and $n_\text{m}$ are the number of quanta in each of the three mechanical modes. Thus, in the absence of other dissipative processes a simple rate equation for the population of the mechanical cavity mode is given by

\begin{equation}
\dot{n}_\text{m} = \Gamma_\text{+}-\Gamma_\text{-} = -A(n_\text{2}-n_\text{1}) n_\text{m} +A (n_\text{2}+1) n_\text{1},
\end{equation}

\noindent which has the same form as the equation for a harmonic oscillator interacting with a bath of occupation $n_\text{p}$ with a coupling rate $\gamma_\text{p}$, where

\begin{equation}
n_\text{p} = \frac{n_\text{2} (n_\text{1}+1)}{n_\text{2}-n_\text{1}}, \quad \gamma_\text{p} = A(n_\text{2}-n_\text{1}).
\end{equation}

\noindent If the two high-frequency modes are both in equilibrium with each other at some elevated temperature $T_\text{p}$, such that $n_\text{i} = (\exp{\left(\frac{\hbar \omega_\text{i}}{k_\text{B}T_\text{p}}\right)}-1)^{-1}$, it is easy to show that $n_\text{p}$ is simply given by the Bose-Einstein occupation factor for the mechanical mode at temperature $T_\text{p}$ ($n_\text{p} = (\exp{\left(\frac{\hbar \omega_\text{m}}{k_\text{B}T_\text{p}}\right)}-1)^{-1}$). We can also see that the scattering rate $\gamma_\text{p}$ will depend on $T_\text{p}$ through the temperature dependence of the population difference $n_\text{2}-n_\text{1}$. This dependence can be approximately linear or exponential in $T_\text{p}$, depending on the value of $(\hbar \omega_\text{2})/(k_\text{B}T_\text{p})$. 

More realistically, the optical absorption process will populate a range of high-frequency phonon modes above some cutoff frequency ($\omega_\text{c}$), which can contribute to the heating. On the assumption that these modes come into equilibrium with each other at some elevated temperature $T_\text{p}$, we can easily show that that the expression for $n_\text{p}$ is unchanged, and the effective bath occupancy $n_\text{p}$ will be given by the above expression. The scattering rate will now be given by

\begin{equation}
\gammapO = \int_{\omega_\text{c}}^{\infty} d\omega \rho(\omega) A(\omega,\omegamO) (n(\omega,\Tbathp)-n(\omega+\omegamO,\Tbathp),
\end{equation}

\noindent where $\rho(\omega)$ is the density of modes at frequency $\omega$, the matrix element $A$ now can depend explicitly on frequency, and $n(\omega,T)$ is just the Bose-Einstein occupation at frequency $\omega$ and temperature $T$. In general, we can obtain limited analytical results if we assume that the product of the density of states and the matrix element obeys some power law as a function of frequency, $\rho(\omega) A(\omega,\omegamO) \propto \omega^a$. For example, under a simple continuum elastic model the product would be given by $\rho(\omega) A(\omega,\omegamO) \propto \omegamO (\omegamO+\omega) \omega^3$~\cite{Srivastava1990}. Making a simple change of variables $x = \frac{\hbar \omega}{k_\text{B} \Tbathp}$, and assuming $\omegamO \ll \omega_\text{c}$, we then arrive at the approximate relation

\begin{equation}
\gammapO \propto \omegamO \Tbathp^{a+1} \int_{x_\text{c}}^{\infty} dx \frac{x^a e^x}{(e^x-1)^2}; \quad x_\text{c} = \frac{\hbar \omega_\text{c}}{k_\text{B} \Tbathp}.
\end{equation}

Considering first the limiting case $x_\text{c} \gg 1$, we note that the integral can be approximated by

\begin{equation}
\int_{x_\text{c}}^{\infty} dx \frac{x^a e^x}{(e^x-1)^2} \approx \int_{x_\text{c}}^{\infty} dx \frac{x^a}{e^x} = \Gamma(a+1,x_c),
\end{equation}

\noindent where $\Gamma(\alpha,z)$ is the upper incomplete Gamma function. For real values of $z$, this function has the asymptotic behavior $\Gamma(\alpha,z) \rightarrow z^{\alpha-1} e^{-z}$ as $|z| \rightarrow \infty$. This leads to the approximate scaling law

\begin{equation}
\gammapO \propto \Tbathp^{a+1} x_\text{c}^a e^{-x_\text{c}} = \Tbathp e^{-\left(\frac{\hbar \omega_\text{c}}{k_\text{B} \Tbathp}\right)},
\end{equation}

\noindent for $\Tbathp \ll \frac{\hbar \omega_\text{c}}{k_\text{B}}$. In the limiting case $x_\text{c} \ll 1$, the lower limit of the integral can be extended to 0, and a simple integration by parts shows that, for $a > 1$,

\begin{equation}
\int_{0}^{\infty} dx \frac{x^a e^x}{(e^x-1)^2} = a \int_{0}^{\infty} dx \frac{x^{a-1}}{e^x-1} = a \Gamma(a) \zeta(a),
\end{equation}

\noindent where $\Gamma(a)$ is the gamma function and $\zeta(a)$ is the Riemann zeta function. As this is a simple constant, we get the scaling law $\gammapO \propto \Tbathp^{a+1}$, when $\Tbathp \gg \frac{\hbar \omega_\text{c}}{k_\text{B}}$ and $a > 1$. 

\begin{figure*}[t]
\begin{center}
\includegraphics[width=1.75\columnwidth]{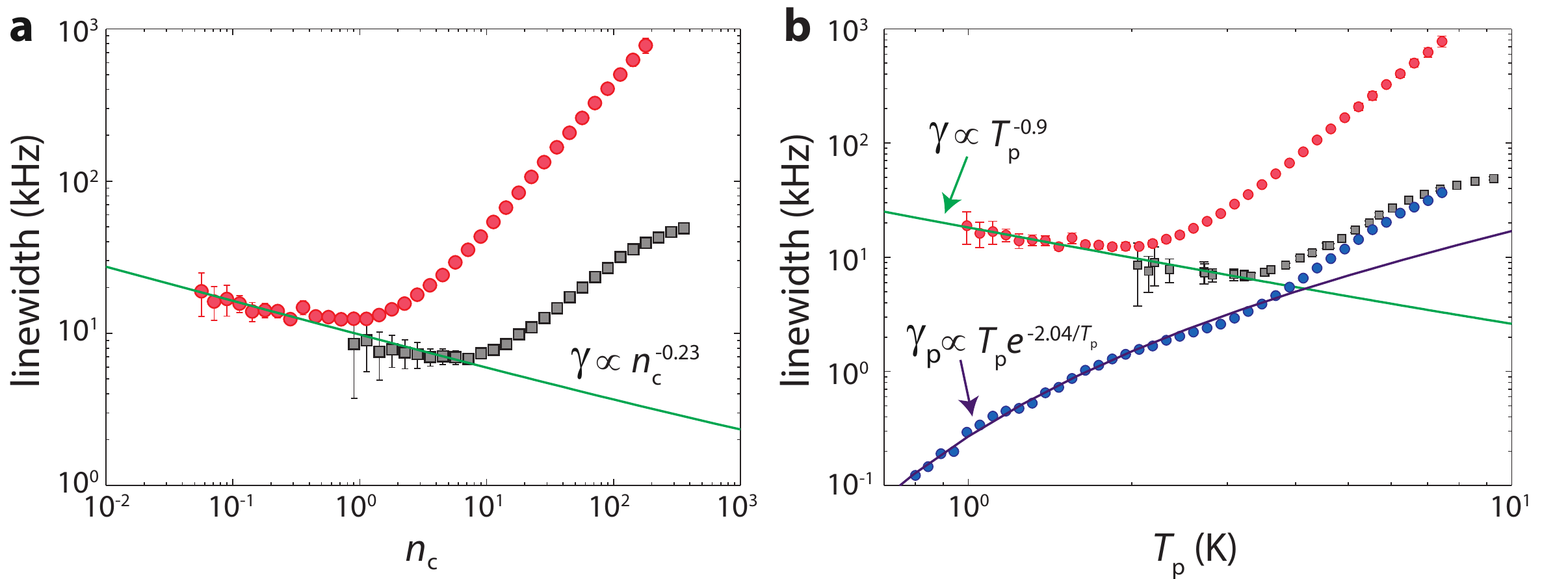}
\caption{\textbf{$\Tf = 10~$mK mechanical linewidth measurements.} \textbf{a}, Mechanical linewidth measured at $\Tf=10~$mK with a red-detuned (red circles) and resonant (black squares) probe versus $\ncavO$, with a power-law fit (solid green line) to the region of both data sets in which frequency jitter is dominant. \textbf{b}, Mechanical linewidth versus optical-absorption-driven temperature $\Tbathp$ as determined from the resonant heating measurement of Fig.~\ref{fig:resonant}a of the main text. The thermalization rate $\gammapO$ extracted from the occupation model is shown as blue circles. A power-law fit corresponding to $\Tbathp^{-0.9}$ of the time-averaged measured linewidth is shown as a solid green curve, showing good correspondence where frequency jitter is dominant. A fit to $\gammapO(\Tbathp)$ at the low temperature end of the data is shown as a purple curve.} \label{Sfig:diffusion}
\end{center} 
\end{figure*}

In Figure~\ref{Sfig:diffusion}b the low temperature end of the inferred $\gammapO(\Tbathp)$ data is seen to fit well to a curve $\propto \Tbathp \exp{\left(-T_{\text{c}}/\Tbathp \right)}$, where the cut-off frequency is $T_{\text{c}} \approx 2$~K ($\omega_{c}/2\pi \approx 35$~GHz).  Although this simple model does not capture all of the features of the measured $\gammapO(\Tbathp)$ curve (in particular, the kink at $\Tbathp \approx 3$~K), it does show that the damping rate due to 3-phonon mixing can vary substantially as a function of $\Tbathp$, particularly at low temperatures in the vicinity of $\Tbathp \lesssim \frac{\hbar \omega_\text{c}}{k_\text{B}}$.

\section{Temperature dependence of frequency noise}
\label{sec:appG}

At sub-kelvin fridge temperatures, frequency noise of the 'breathing' mode resonance is seen to dominate the time-averaged measured linewidth for low optical probe powers (Fig.~\ref{Sfig:diffusion}a). In the case of the red-detuned probe, frequency noise is observed for $\ncavO < 1$ before optomechanical damping kicks in, while the on-resonance measurement shows frequency noise for $\ncavO < 10$ before the intrinsic energy damping ($\gammapO$) begins to dominate. Combining the sections of both data sets which are dominated by frequency noise, a power law fit shows linewidth scaling as $\ncavO^{-0.23}$.  By comparing to the on-resonance measured mode occupancy (Fig.~\ref{fig:resonant}a in main text), the frequency noise can also be plotted with respect to the optical absorption driven temperature $\Tbathp$ as shown in Fig.~\ref{Sfig:diffusion}b. A power law fit of the time-averaged linewidth in the frequency jitter dominated regime, shows a frequency jitter scaling $\propto \Tbathp^{-0.9}$. 

Similar inverse temperature scaling of frequency noise has been observed in superconducting microwave resonators ~\cite{GaoPhD}, with substantial evidence indicating the source to be fluctuations in two-level tunneling states (TLS) of near-field amorphous materials~\cite{Phillips1972,Anderson1972}. These TLS also couple to phonons, contributing to not only the dielectric properties of the material, but also the elastic properties. As such, the microwave mechanical modes of the devices studied here can be expected to couple to TLS in a similar fashion as for the microwave electromagnetic resonators.  The decrease in frequency noise with temperature can then be attributed to the thermal excitation and saturation of the TLS.  The presence of TLS in the Si devices of this work likely stems from the formation of a native oxide on the Si surfaces of the patterned nanobeam.  Careful measures, involving a final wet etch in HF acid~\cite{Borselli2007}, were made to passivate the Si surfaces of the nanobeam in order to reduce the optical absorption from sub-bandgap surface electronic states.  This surface cleaning should have been effective in removing the surface oxide as well, however, a one hour procedure was required to load the sample into the dilution refrigerator, during which time a native oxide would at least partially reform on the Si surfaces of the device.  This native oxide is likely to be the source of the TLS.  Future work will look to substantially reduce the device loading time after HF surface cleaning, hopefully reducing both the optical absorption and TLS density.   

\end{document}